\documentclass[12pt,onecolumn,draftclsnofoot]{IEEEtran}

\usepackage{cite}
\usepackage{psfrag}
\usepackage[pdf]{graphicx}
\usepackage[latin1]{inputenc}
\usepackage[T1]{fontenc}
\usepackage{amsmath,amsfonts,amsbsy,amssymb}
\usepackage{mathabx}
\usepackage{mathrsfs}
\usepackage[nolist]{acronym}
\usepackage{tabularx}
\usepackage{amssymb}
\usepackage{amsmath}
\usepackage{graphicx}
\usepackage[latin1]{inputenc}
\usepackage{graphicx}
\usepackage{cite}
\usepackage{multirow}
\usepackage{wasysym}
\usepackage{multirow}
\usepackage{float}
\usepackage{color}

\newtheorem{definition}{Definition}
\newtheorem{proposition}{Proposition}
\newtheorem{lemma}{Lemma}

\newtheorem{corollary}{Corollary}
\newtheorem{remark}{Remark}

\newcommand{\new}[1]{{\color{black}#1}}

\begin{document}

\title{Maximizing the Link Throughput between Smart meters and Aggregators as Secondary Users under Power and Outage Constraints}
\author{
	\IEEEauthorblockN{Pedro H. J. Nardelli, Mauricio de Castro Tomé,\\ Hirley Alves, Carlos H. M. de Lima and Matti Latva-aho} 
	\thanks{P. H. J. Nardelli, M. C. Tomé, H. Alves and M. Latva-aho are with the Centre for Wireless Communications (CWC), University of Oulu, Finland. Contact: nardelli@ee.oulu.fi. 
	C. H. M. de Lima is with São Paulo State University (UNESP), São João da Boa Vista, Brazil.
	This work is partly funded by Finnish Academy and CNPq/Brazil (n.490235/2012-3) as part of the joint project SUSTAIN, by Strategic Research Council/Aka BC-DC project (n.292854), and  by the European Commission through the P2P-SmarTest project (n.646469).}
}
\maketitle
\vspace{-1.6cm}
%
\begin{abstract}
\vspace{-0.2cm}
This paper assesses the communication link from smart meters to aggregators as (unlicensed) secondary users that transmit their data over the (licensed) primary uplink channel.
The proposed scenario assumes: (i) meters' and aggregators' positions are fixed so highly directional antennas are employed, (ii) secondary users transmit with limited power in relation to the primary, (iii)  meters' transmissions are coordinated to avoid packet collisions, and (iv) the secondary links' robustness is guaranteed by an outage constraint.
Under these assumptions, the interference caused by secondary users in both primary (base-stations) and other secondary users can be neglected.
As unlicensed users, however, meter-aggregator links do experience interference from the mobile users of the primary network, whose positions and traffic activity are unknown.
To cope with this uncertainty, we model the mobile users spatial distribution as a Poisson point process.
We then  derive a closed-form solution for the maximum achievable throughput with respect to a reference secondary link subject to transmit power and outage constraints.
Our numerical results illustrate the effects of such constraints on the optimal throughput, evincing that more frequent outage events improve the system performance in the scenario under study.
We also show that relatively high outage probabilities have little effect on the reconstruction of the average power demand curve that is transmitted from the smart meter to the aggregator.
\end{abstract}
\vspace{-0.5cm}
\begin{keywords}
\vspace{-0.5cm}
cognitive network, Poisson point process, smart grid communication, spectrum sharing
\end{keywords}
%

\section{Introduction}
Over the last few years, cognitive radios have appeared as the solution for more effective use of the frequency spectrum \cite{akyildiz2006next, Cormio2009}.
Following the concept proposed by Haykin \cite{Haykin2005}, the radio nodes should understand their environment to establish a wireless network ``(...) with two primary objectives in mind: highly reliable communication whenever and wherever needed; efficient utilization of the radio spectrum.''
Owing to their cognitive ability, radios would be then capable of sensing the environment to decide on their transmission.

An interesting approach to the cognitive radio concept is the so-called spectrum sharing \cite{Peha2008}, where unlicensed -- secondary -- users want to transmit some information without disturbing the licensed -- primary -- users over the same frequency band.
Secondary users then need to sense the spectrum usage to decide about their transmissions.
The transmission occurs if the channel is sensed free, which depends on the primary user activity \cite{saleem2014primary}.
Otherwise, the secondary user either searches for a different band or postpones its transmission.

Although the basic idea is simple, both analysis and implementation are challenging due to interactive dynamics of the decision-making procedures \cite{wang2010game}.
The solution gets even harder when we consider different possible applications and their specific requirements, yielding no universal solution \cite{Ahmad2015}.
Notwithstanding, the idea of spectrum sharing is now widespread in both new generations of cellular system (allowing for co-existence of macro- and small-cells as in \cite{de2012coordination}) and different sensor applications (e.g. \cite{Xu2014,kelly2013towards,bradai2015emcos}).

In this paper, we focus on the specific application of  spectrum sharing in the deployment of part of the communication network in the modern electric power grids \cite{bush2014smart,nardelli2014models} -- the so-called smart grids. 
As one would expect, the different applications of the electric power grid have different requirements from the communication network perspective \cite{Yan2013}.
\new{For instance, control operations in the high-voltage grid must be close to real-time (scale of milliseconds) so the communication link must have both extremely low latency and very high reliability  (greater than $99.9\%$).
In contrast, remote reading of meters would allow for less stringent communication requirements (latency of minutes and a reliability of $98\%$).}
As a consequence, the communication network design should be fine-tuned with the specific power grid application and its needs.
An informative survey about the different applications and requirements is found in \cite{kuzlu2014communication}.

Looking specifically at how to implement cognitive radio for smart grid applications, the authors in \cite{qiu2011cognitive} presented an interesting hardware study case using software defined radio in a micro-grid testbed.
In \cite{shah2013cross}, a Lyapunov-drift framework was proposed to differentiate traffic priorities and then use the cognitive radio strategy called ``dynamic spectrum access'' to improve the communications in smart grids.
Additionally, \cite{khan2015cognitive} provides a recent survey on how to combine cognitive radio approach into smart grid scenarios.

\new{In our case, references \cite{yu2011cognitive} and \cite{gungor2012cognitive} introduce the idea of employing a spectrum sharing scheme within households where home appliances use unlicensed channel bands to build a home area network (HAN).
In neighborhood area networks (NANs), different scenarios may be considered for the smart meters: (i) licensed users within the cellular systems directly connected with the distribution operator \cite{nielsen2015can}, (ii) users of unlicensed bands of unused TV frequencies (TV white spaces) \cite{zhang2012cognitive}, (iii) secondary users of licensed band  \cite{yu2011cognitive,gungor2012cognitive} and (iv) hybrid licensed-unlicensed users \cite{gungor2012cognitive}.
When using solutions (i) and (iv), the smart grid elements are subscribers (licensed users) of cellular networks.
In this case, although quality of service shall be guaranteed, they may involve prohibitive high costs and a dramatic increase in data traffic \cite{nielsen2015can}, making them not so attractive for electricity providers if compared to (ii) and (iii).
And yet, the latter solutions will have a strong dependence on specific country legislation.}

\new{Despite such drawbacks, these solutions are interesting and related to ours.
We believe, however, that the analysis provided by the aforementioned works are focused on network-layer considerations without a dedicated performance study involving specific topological considerations as node positions and mobility.
Herein, we attempt to cope with such limitations by studying a novel application of spectrum sharing technique for distribution grids considering that some elements are fixed and others are mobile.}

In specific terms, a spectrum sharing scheme (e.g \cite{Peha2008}) will be assumed such that secondary users transmit in the up-link channel of the cellular system.
As meters and aggregators are generally static nodes, their communication link may be built using directional antennas (e.g. \cite{Wildman2014}) while respecting a transmit power constraint.
In this way, the harmful effects caused by the secondary user transmissions on the primary users, as well as on other secondary users, due to co-channel interference are limited.
By properly designing the antenna beamforming and setting the power constraint, the probability that the secondary users interfere in the primary transmissions would be low as far as directional antennas with limited power have a restrict radiation pattern.
Besides, by using the up-link channel, the cellular base-stations are disturbed by the secondary transmissions and, due to their more robust reception procedures \cite{Andrews2005}, the interference can be further mitigated.

On the other hand, the secondary users do experience the interference caused by the primary users.
Since the up-link is employed by the former, the interference normally comes from mobile devices, whose positions and traffic are unknown.
To account for such uncertainties, the interferers' spatial distribution and traffic characteristics will be modeled using point process theory \cite{haenggi2012stochastic}.
Thereby, it is possible to derive closed-form expressions for important performance metrics of wireless systems as outage probability and link throughput.
It is worth mentioning that the communication between different smart meters and their respective aggregator is coordinated so they do not interfere to each other.

Our goal in this work is to optimize the link throughput (which is defined as the spectral efficiency, given in bits per second per hertz, times the probability that the packet is successfully decoded by the aggregator) of the secondary link under power (to not affect the primary users) and outage (secondary link reliability) constraints.
Using a similar optimization procedure as in \cite{Nardelli2012b,Nardelli2014,Nardelli2015},  we find the signal-to-interference ratio threshold and the transmit power employed by the secondary link so as to maximize its throughput while respecting the imposed constraints.

Then, we found a closed-form equation for the optimal link throughput as a  non-linear function of the system constraints and the density of interferers, as well as other system parameters.
Interestingly, our results show that frequent outage situations have an unexpected positive effect on the system performance for the scenario of interest (maximum value of outage constraint is $25\%$).
We use actual data from the average power demand of a household (obtained from ``The Reference Energy Disaggregation Data Set'' database \cite{kolter2011redd,REDD}) to show that relatively high outage probabilities do not lead to a poor signal reconstruction if the information is sampled and sent periodically, for example, every $15$ minutes.
All in all, our main contribution and novelty are the theoretical evidences that up-link spectrum sharing is a good candidate for deploying the communication network in the distribution electricity grids.

The rest of this paper is divided as follows.
Section \ref{sec:SystemModel} explains the scenario under analysis, justifying our assumptions and presenting the optimization problem to be solved.
In Section \ref{sec:Max}, we focus on the solution of the optimization problem and illustrate how the system performance changes with the configuration parameters.
Section \ref{sec:signal-rec} shows how the outage events affect the reconstruction of the average power demand curve.
In section \ref{sec:disc}, we discuss some implications of our theoretical results, indicating how they might be used in actual deployments.
Section \ref{sec:Conc} concludes this paper.

\section{System model}
\label{sec:SystemModel}
\new{In this section, we describe the basic assumptions used to build our model and their implications.
The assumptions are stated as follows:
\begin{itemize}
\item \textbf{Assumption 1}: Spectrum-sharing scenario where licensed (primary) and unlicensed (secondary) users share the frequency bands allocated to the up-link channel.
\item \textbf{Assumption 2}: Primary link is established between static cellular base-stations and mobile users. Secondary terminals are smart meters that need to forward data to a given aggregator through the uplink channel and their positions are fixed.
\item \textbf{Assumption 3}: The smart meters transmit with limited power $W_\mathrm{s}$ such that $W_\mathrm{s} \leq W_\mathrm{s, max}$ where $W_\mathrm{s, max}$ is the maximum power allowed for the secondary users (which can be seen as an imposition from the primary network).
\item \textbf{Assumption 4}: Smart meters associated with the same aggregator are able to perfectly coordinate their transmissions using time scheduling.
\end{itemize}
In this case, Assumption 2 indicates the possibility of employing directional antennas in the secondary links as far as their positions are fixed.
Orientation errors as defined in  \cite{Wildman2014} can be then completely avoided when deploying the secondary network, making highly directional antennas worth. 
In its turn, Assumption 3 imposes the maximum range that the signal transmitted by the smart meters can reach.
Putting all together, the radiation pattern generated by the secondary transmission can be seen as a line segment starting in the smart meter, passing through the aggregator and ending in a point related to $W_\mathrm{s, max}$.

Let us now look at the interference related to the proposed spectrum sharing. 
Assumptions 1, 2 and 4 indicate the co-channel interference occurs: (i) from smart meters to cellular base-stations, (ii) from smart meters to aggregators that they are not associated, and (iii) from mobile users to aggregators.
From the implications discussed in the previous paragraph, the cases (i) and (ii) can be neglected by designing the specific locations when the deploying either the secondary or primary networks.
Even if the positions are considered random in two-dimensions, the chance of having a base-station or an aggregator in the line segment related to the smart meter transmitted signal approaches zero, which further indicated that such cases should not be considered.

Hence, only case (iii) is relevant for our analysis.
To evaluate its impact in the system performance, we first need to model uncertainty of the mobile users' positions and traffic activity.}
We assume here a Poisson field of interferers \cite{Cardieri2010} such that the interferer nodes are distributed over an infinite plane following a $2$-dimensional Poisson point process $\Phi$ with density $\lambda$, given in interferers per square-meter.
The wireless channel model employed in this paper consists of two components: one related to the distance-dependent path-loss such that the received power decays with the distance and other related to fast-fading \cite{haenggi2012stochastic}.
The received power at the node of interest can be computed as $g_i
r_i^{-\alpha}$, where $r_i$ is the distance between the reference receiver and the $i\text{th}$ node, $g_i$ is the channel gain between them, and $\alpha>2$ the path-loss exponent.
%

%
%
%

\begin{figure}[!t]%
	\centering
	\includegraphics[width=0.75\columnwidth]{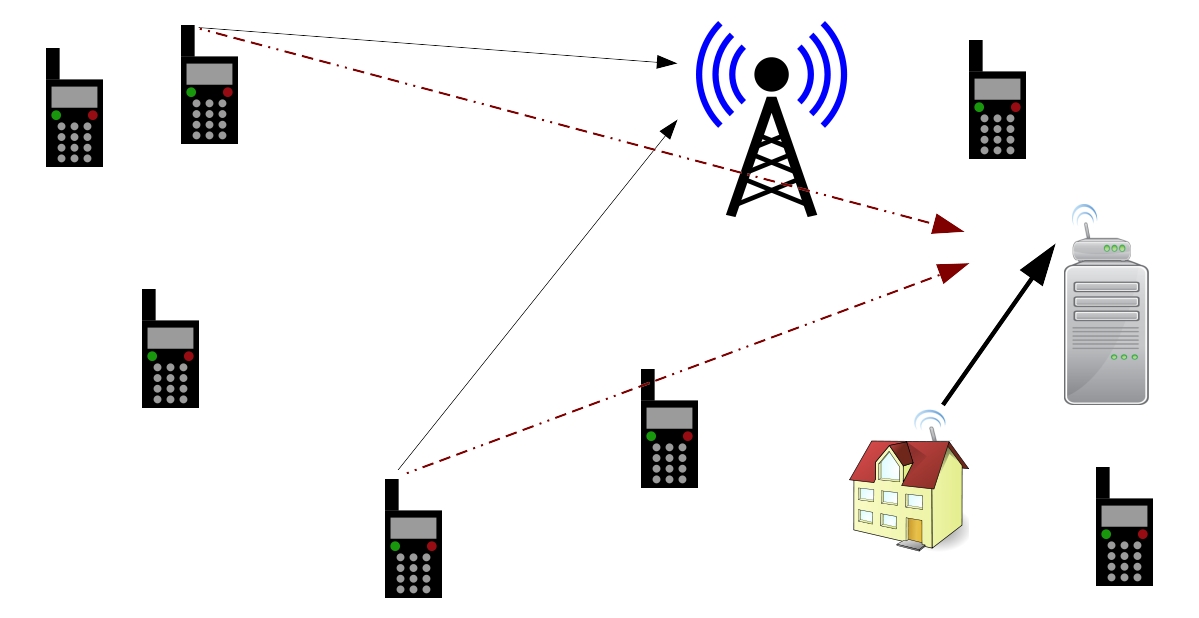}
	%
	\caption{\new{An illustration of the proposed scenario, where primary and secondary users share the up-link channel. The reference smart meter (secondary transmitter) is depicted by the house, the aggregator (secondary receiver) by the the CPU , the handsets are the mobile primary users (interferers to the aggregator) and the big antenna is the cellular base-station. As the smart meter uses directional antenna with limited transmit power (bold arrow), its interference towards the base-station can be ignored. The thin black arrows represent the primary users' desired signal, while the red ones their interference towards the aggregator.}}
	\label{fig-scheme}
	%
\end{figure}

Hereafter, we focus our analysis on a reference smart meter-aggregator link, as shown in Fig.~\ref{fig-scheme}.
During transmissions intervals, we assume that the interferers' positions and the channel gains do not change.
We also consider an interference-limited scenario wherein the noise effects can be neglected. 
As pointed in \cite{Weber2010}, the inclusion of the noise power leads to a more complex analysis without providing any significant qualitative difference. 

If the primary users are equipped with omni-directional antennas and transmit with the same fixed power $W_\mathrm{p}$, the signal-to-interference ratio (SIR) at the aggregator can be computed as \cite{haenggi2012stochastic}:
\begin{equation} 
\label{eq_SIR}
	\textrm{SIR}_0 = \dfrac{W_\mathrm{s} g_{0}  r_0^{-\alpha}}{W_\mathrm{p} \; \underset{i \in \Phi}{\displaystyle \sum}  g_{i}  r_i^{-\alpha}},
\end{equation}
where the index $0$ denotes the reference link.

We assume that the reference link employs both point-to-point Gaussian codes and the interference-as-noise decoding rule \cite{Baccelli2011,Nardelli2015} so that a spectral efficiency of $\log_2(1 + \beta)$ in bits/s/Hz is achieved only if the SIR is greater than $\beta$.
In this case, the probability $P_\mathrm{suc}$ that a packet is successfully decoded by the aggregator is the probability that $\textrm{SIR}_0 > \beta$.
Then, an outage event happens with probability $1 - P_\mathrm{suc}$.
In our scenario, retransmissions are not allowed so that the information contained in packets received in outage is lost.

To compute  $P_\mathrm{suc}$, we assume quasi-static channel gains (squared envelops) $g$ that are independent and identically distributed exponential random variables (Rayleigh fading), and also a dynamic topology where interferers' positions change every transmission interval.
Therefore, every transmission attempt can be viewed as a different realization of the point processes $\Phi$ and the channel gains $g$.
We consider here that the distance between the reference meter and aggregator is known and has a fixed value $r_0=d$.
Finally, the success probability is computed as \cite{haenggi2012stochastic}:
\begin{equation} 
\label{eq_Psuc}
	P_\mathrm{suc}= e^{-  \lambda \kappa \pi d^2 \beta^{2/\alpha}
	\left(\frac{W_\mathrm{p}}{W_\mathrm{s}}\right)^{2/\alpha}},
\end{equation}
where $\kappa = \Gamma(1 + 2/\alpha)\Gamma(1 - 2/\alpha)$ with $\Gamma(\cdot)$ being the Gamma function.
It is also worth noting that the primary users might employ different power control strategies, but it will not affect the qualitiative results of our analysis due to the stochastic geometry approach employed (e.g. \cite{weber2007effect}).

We are now ready to define the performance metric and carry out the optimization problem under consideration.
\begin{definition}[Link throughput]
The throughput $T$ of the reference link using the system model described in this section is defined as:
\begin{equation}
\label{eq:throuput}
	T = \log(1 + \beta) \; P_\mathrm{suc} = \log_2(1 + \beta) \; e^{-  \lambda \kappa \pi d^2 \beta^{2/\alpha}
		\left(\frac{W_\mathrm{p}}{W_\mathrm{s}}\right)^{2/\alpha}}.
\end{equation}
\end{definition}

Our goal in this paper is to find the setting of parameters for the secondary users to maximize their link throughput $T$ while respecting the imposed power limit and outage constraints.
In our case, the variables in hand are the coding rate $\beta$ and the transmit power $W_\mathrm{s}$ of the reference link. 
Mathematically, we have the following problem:
\begin{equation}
\label{eq:opt-prob}
	\begin{aligned}
		&  \underset{(\beta, W_\mathrm{s})}{\max} & & \log_2(1 + \beta) \; e^{-  \lambda \kappa \pi d^2 \beta^{2/\alpha}
				\left(\frac{W_\mathrm{p}}{W_\mathrm{s}}\right)^{2/\alpha}} \\
		&  \text{s.t. } 	& & W_\mathrm{s} \leq W_\mathrm{max} \\
		& 					& & 1 - P_\mathrm{suc} \leq \epsilon
	\end{aligned},
	%
\end{equation}
where $\epsilon$ is the maximum acceptable outage probability, reflecting the  reference link robustness.

\vspace{-2ex}
\section{Maximum throughput under power and outage constraints}
\label{sec:Max}
In this section, we solve the optimization problem previously stated.
We then provide some numerical results to illustrate how the constraints imposed to our smart meter-aggregator reference link will affect the maximum achievable throughput.
Before we start, we still need to present a Lemma that tells us how the throughput behaves as a function of the secondary transmit power $W_\mathrm{s}$ and the SIR threshold $\beta$ when no constraint is considered.

\begin{lemma}
\label{lemma:function-analysis}
Let us consider the throughput equation, given by \eqref{eq:throuput}, as a function of the variables $W_\mathrm{s}>0$ and $\beta>0$, i.e. $T = f\left(W_\mathrm{s}, \beta\right)$.
\new{The function $f$ is monotonically crescent in respect to $W_\mathrm{s}$, and it is concave in respect to $\beta$ if $\partial^2 T/\partial \beta^2 < 0$}.
\end{lemma}
\begin{IEEEproof}[Outline of proof]
The proof of this Lemma is straightforward from the analysis of the function in terms of the (strictly positive) variables $W_\mathrm{s}$ and $\beta$, and knowing that the function $T$ is twice differentiable in terms of $\beta$.
Any increase in $W_\mathrm{s}$ leads to an increase in the exponential term of \eqref{eq:throuput} and then in $T$.
Increasing $\beta$, on the other hand, has a two-fold effect: it increases the logarithmic term while decreases the exponential one.
\new{The function $T$, however, is not always concave in relation to $\beta>0$; nevertheless, in the region that $\partial^2 T/\partial \beta^2 < 0$, $T$ is concave.}
\end{IEEEproof}

\new{Since $T$ is concave for some values of $\beta$, we may try to find the value of $\beta$ that leads to the maximum $T$.
This is shown in the next Lemma.}

\begin{lemma}
\label{lemma:optimal-beta}
\new{Let $\beta^*_\mathrm{un}$ denote the value of $\beta$ that maximizes $T$ assuming that $\beta^*_\mathrm{un}$ is in the region where $\partial^2 T/\partial \beta^2 < 0$.
Then, $\beta^*_\mathrm{un}$  is the solution of the following (transcendental) equation:}
\begin{equation}
\label{eq:beta-opt-uncons}
\alpha \beta = k \; \beta^{2/\alpha} \; (1+\beta) \; \ln(1+\beta),
\end{equation}
where $k = 2 \lambda \kappa \pi d^2 \left(W_\mathrm{p}/W_\mathrm{s}\right)^{2/\alpha}$.
\end{lemma}
\begin{IEEEproof}[Outline of proof]
\new{Let us first consider that $\beta^*_\mathrm{un}$ is in the region where the inequality $\partial^2 T/\partial \beta^2 < 0$ holds.
Then, from Lemma \ref{lemma:function-analysis}, the value of $\beta$ that maximizes $T$ is the solution of the derivative equation $\partial T / \partial \beta = 0$.
By doing so in \eqref{eq:throuput}, we end up in \eqref{eq:beta-opt-uncons}, which has no closed-form solution.  
If a solution does not exist, then our initial assumption the inequality $\partial^2 T/\partial \beta^2 < 0$ does not hold and $\beta^*_\mathrm{un}$ cannot be obtained.}
\end{IEEEproof}

\new{Nonetheless, in the cases of interest, \eqref{eq:beta-opt-uncons} has a solution and therefore $\beta^*_\mathrm{un}$ is in the region where $\partial^2 T/\partial \beta^2 < 0$ as we shall see next.}
It is noteworthy that even though \eqref{eq:beta-opt-uncons} does not have a closed-form solution, it can be easily evaluated numerically through standard mathematical software as, for instance, \cite{fsolve-scipy}.

\begin{figure}[!t]
	\centering
	\includegraphics[width=\columnwidth]{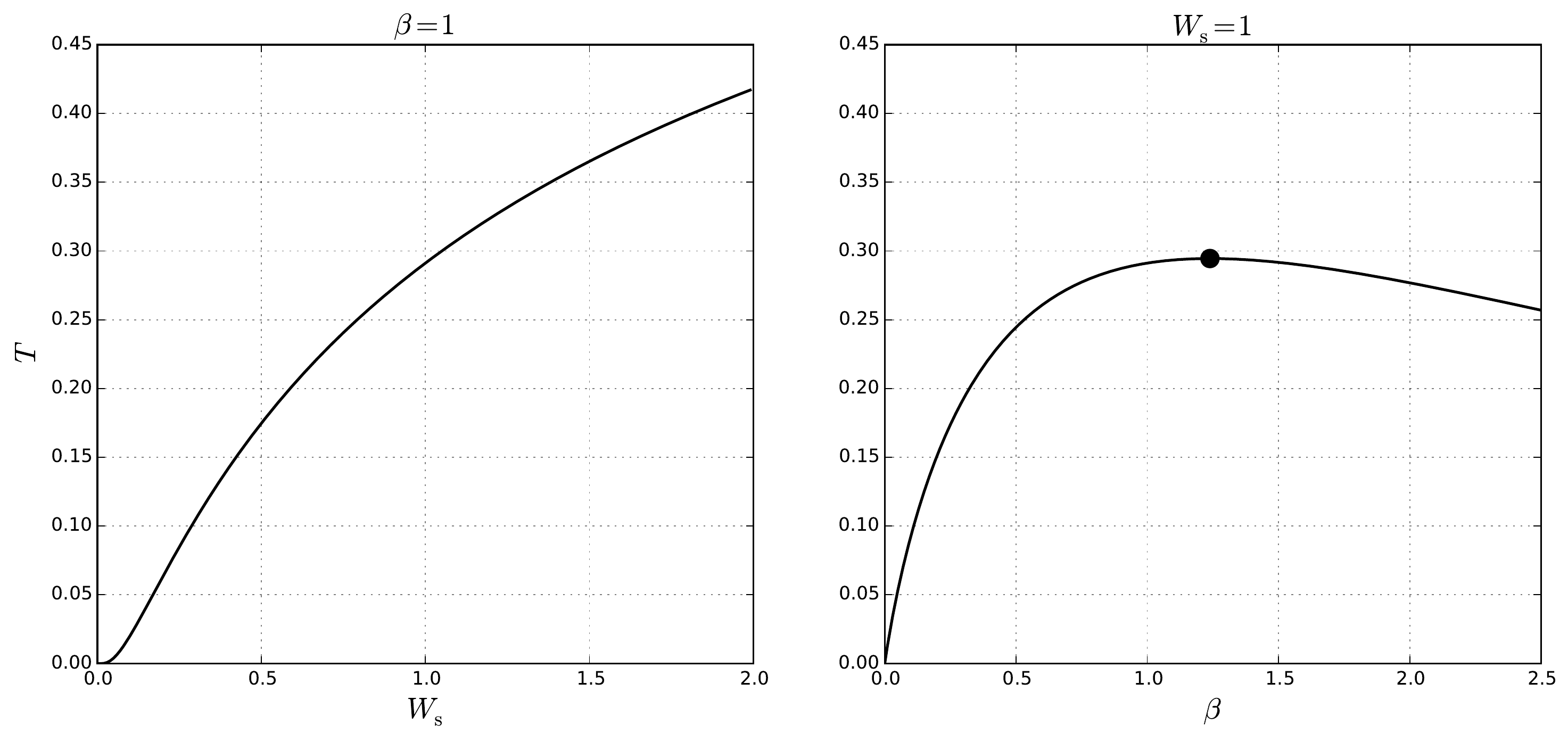}
	\caption{Link throughput $T$ given in \eqref{eq:throuput} as a function of the transmit power $W_\mathrm{s}$ (left) and the SIR threshold $\beta$ (right) for $\alpha = 4$, $d = 1$, $\lambda = 0.25$ and $W_\mathrm{p} = 1$. The black dot in the right plot is the optimal operating point predicted by Lemma \ref{lemma:optimal-beta}.}
	\label{fig:uncons-exemp}
\end{figure}

Fig. \ref{fig:uncons-exemp} exemplifies the behavior of the link throughput $T$ with the transmit power $W_\mathrm{s}$ and the threshold $\beta$.
The curve behaves as predicted by Lemmas \ref{lemma:function-analysis} and \ref{lemma:optimal-beta}. 
Intuitively, when all parameters are kept the same, it is advantageous for the smart meter to increase its transmit power so that the SIR experienced by the aggregetor tends to increase.
This behavior, although individually optimal, is not good for the other links as transmitting with more power would increase the interference level throughout the network (refer to \cite{Nardelli2014} for a deeper assessment of an  optimization problem wherein the optimal individual solution can be socially harmful).
For this reason, the power constraint that our problem assumes is needed, as we will discuss later.

When analyzing the effects of the threshold $\beta$, there is a trade-off involved.
An increase of $\beta$ leads to a more efficient transmission where more bits/s/Hz can be transmitted in the same message.
This gain, however, comes at expense of more frequent outage events, which in turn decreases the link throughput.
From Lemma \ref{lemma:optimal-beta}, the value of $\beta$ that leads to the optimal operating point can be found by numerically solving \eqref{eq:beta-opt-uncons}.
In our case, we use the function $\mathrm{fsolve}(\;)$ from the Python library SciPy \cite{fsolve-scipy}.

We are now almost ready to present the main result of this paper, which is the solution of the constrained optimization problem presented in the previous section.
But first, we still need to state a lemma about the relation between the optimal values of the constrained and unconstrained SIR thresholds $\beta$.

\begin{lemma}
\label{lemma:optimal-beta-vs-ep}
Let $\beta^*$ denote  the value of $\beta$ that leads to the maximum constrained throughput given in \eqref{eq:opt-prob} and $\beta^*_\mathrm{un}$, given in Lemma \ref{lemma:optimal-beta}, denote the value that optimizes the unconstrained throughput.
Then:
\begin{equation}
\label{eq:opt-beta-vs-ep}
 1 - e^{-\frac{\alpha \beta^*_\mathrm{un}}{2 (1+\beta^*_\mathrm{un})\ln(1+\beta^*_\mathrm{un})}} > \epsilon \; \; \Longrightarrow \; \; \beta^* < \beta^*_\mathrm{un}.
\end{equation}
\end{lemma}
\begin{IEEEproof}[Proof]
In the optimal unconstrained operating point $\beta^*_\mathrm{un}$ given in Lemma \ref{lemma:optimal-beta}, the following equality must hold:
\begin{equation}
\label{eq:beta-opt-uncons-2}
\beta^{2/\alpha} = \dfrac{\alpha \beta}{k  (1+\beta) \ln(1+\beta)}.
\end{equation}

Then, although we cannot analytically compute the actual value of $\beta^*_\mathrm{un}$, we do know that the outage probability $1-P_\mathrm{suc}$, given in \eqref{eq_Psuc}, related to it will be given by: $1 - e^{-\frac{\alpha \beta^*_\mathrm{un}}{2 (1+\beta^*_\mathrm{un})\ln(1+\beta^*_\mathrm{un})}}$, which is a monotonically decreasing function of $\beta^*_\mathrm{un}$.

In this case, if that probability is smaller than the constraint $\epsilon$, then the constrained optimal threshold is the unconstrained one, i.e. $\beta^* = \beta^*_\mathrm{un}$.
Otherwise, the optimal constrained threshold $\beta^*$ must be smaller than $\beta^*_\mathrm{un}$ since the outage probability is a decreasing function of $\beta^*_\mathrm{un}$.
\end{IEEEproof}

\begin{proposition}
\label{prop:optimization-solution}
Let us assume that the pair $(W_\mathrm{s}^*, \beta^*)$ is the solution of the optimization problem given by \eqref{eq:opt-prob}.
If  $1 - e^{-\frac{\alpha \beta^*_\mathrm{un}}{2 (1+\beta^*_\mathrm{un})\log(1+\beta^*_\mathrm{un})}} > \epsilon$, the pair $(W_\mathrm{s}^*, \beta^*)$ is computed as:
\begin{eqnarray}
\label{eq:opt-W-s}
	  & W_\mathrm{s}^*  = W_\mathrm{max},\\ 
\vspace{3ex}
\label{eq:opt-beta}
	&\beta^*	 = \dfrac{W_\mathrm{max}}{W_\mathrm{p}} \left( - \dfrac{\ln(1 - \epsilon)}{\lambda \kappa \pi d^2 }\right)^{\alpha/2}.
\end{eqnarray}

The optimal throughput $T^*$ is then:
\begin{equation}
\label{eq:opt-T}
	T^*= (1 - \epsilon) \times \log_2\left(1 + \dfrac{W_\mathrm{max}}{W_\mathrm{p}} \left( - \dfrac{\ln(1 - \epsilon)}{\lambda \kappa \pi d^2 }\right)^{\alpha/2}\right).
\end{equation}
\end{proposition}

\begin{IEEEproof}
Let us start by considering the variable $W_\mathrm{s}$.
From Lemma \ref{lemma:function-analysis}, we know that the throughput is a monotonically increasing function of $W_\mathrm{s}$, regardless of $\beta$.
Then, $W_\mathrm{s}$ must assume its highest possible value: $W^*_\mathrm{s}=W_\mathrm{max}$.
To find  $\beta^*$, we first use the assumption that $1 - e^{-\frac{\alpha \beta^*_\mathrm{un}}{2 (1+\beta^*_\mathrm{un})\ln(1+\beta^*_\mathrm{un})}} > \epsilon$.
From Lemma \ref{lemma:optimal-beta-vs-ep}, the inequality $\beta^* < \beta^*_\mathrm{un}$ holds.
Then, we use the fact that $T$ is also a monotonically crescent function of $\beta$ in the range $0<\beta < \beta^*_\mathrm{un}$ so that $\beta^*$ should be the highest value that satisfies the inequality $1 - P_\mathrm{suc} \leq \epsilon$.
Manipulating the constraint by assuming $W^*_\mathrm{s}=W_\mathrm{max}$, we obtain $\beta \leq \frac{W_\mathrm{max}}{W_\mathrm{p}} \left( - \frac{\ln(1 - \epsilon)}{\lambda \kappa \pi d^2 }\right)^{\alpha/2}$.
In this case, equality gives $\beta^*$.
\end{IEEEproof}

\begin{remark}
\label{rem-beta-opt}
This result is only valid if the initial assumption  $1 - e^{-\frac{\alpha \beta^*_\mathrm{un}}{2 (1+\beta^*_\mathrm{un})\ln(1+\beta^*_\mathrm{un})}} > \epsilon$ holds, which is true for the cases of interest, namely $\alpha \in(2,6]$ and $\epsilon \in (0,0.25)$. Notice that $\alpha \in(2,6]$ comprises indoor and outdoor scenarios as well as rural and urban areas \cite{Goldsmith2005}. 
For example, when $\alpha = 4$ and $\beta = 1.24$ (the value of $\beta^*_\mathrm{un}$ in the example used in Fig. \ref{fig:uncons-exemp}), the outage probability is $0.75$.
If that inequality does not hold, the optimal solution is $\beta^* = \beta^*_\mathrm{un}$ and the optimal power $W^*_\mathrm{s}$ should be computed accordingly.
\new{In other words, the value of $\beta^*$ is kept fixed, while the optimal power $W^*_\mathrm{s}$ is the variable used to optimize the link throughput.}
\end{remark}

\begin{corollary}
\label{col:T-approx}
The optimal throughput $T^*$ can be approximated by:
\begin{equation}
\label{eq:opt-T-approx}
	T^* \approx \dfrac{(1 - \epsilon) \;W_\mathrm{max}}{\ln(2) \; W_\mathrm{p}} \left( - \dfrac{\ln(1 - \epsilon)}{\lambda \kappa \pi d^2 }\right)^{\alpha/2}, 
\end{equation}
when $ \frac{W_\mathrm{max}}{W_\mathrm{p}} \left( - \frac{\ln(1 - \epsilon)}{\lambda \kappa \pi d^2 }\right)^{\alpha/2}$ is small.
\end{corollary}

\begin{remark}
\label{rem-T-approx}
Corollary \ref{col:T-approx} holds for the cases under study.
For example, when the system variables are set as follows: $\lambda=0.25$, $\alpha = 4$, $d = 1$, $W_\mathrm{p} = 1$, $W_\mathrm{p} = 0.5$ and $\epsilon=0.05$, the term is $ \frac{W_\mathrm{max}}{W_\mathrm{p}} \left( - \frac{\ln(1 - \epsilon)}{\lambda \kappa \pi d^2 }\right)^{\alpha/2} = 0.00086$, therefore the approximation $\ln(1+x) \approx x$ works well.
\end{remark}

\begin{figure}[!t]
	\centering
	\includegraphics[width=0.7\columnwidth]{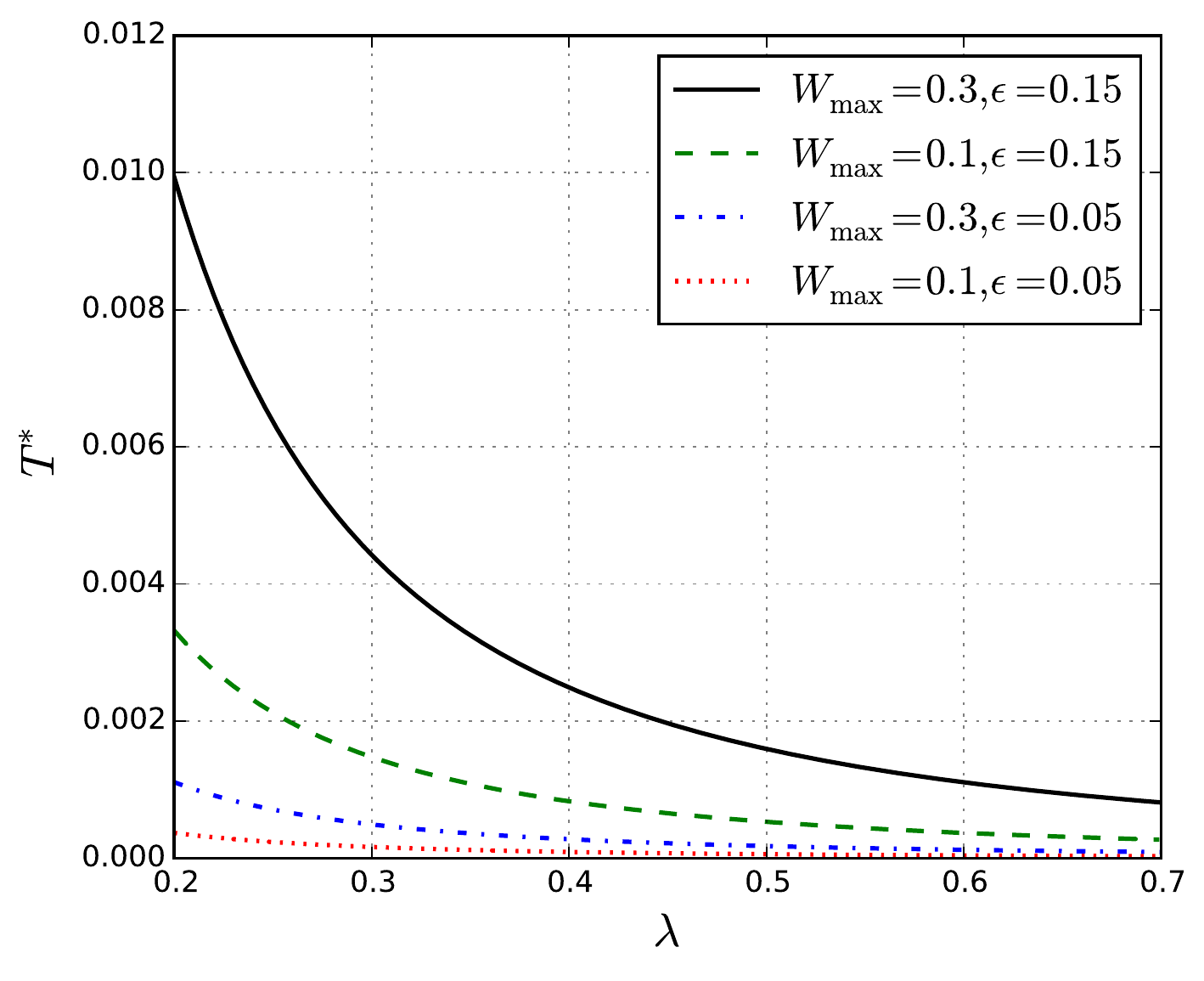}
	\caption{Maximum link throughput $T^*$ given in \eqref{eq:opt-T} as a function of the density of active mobile primary users (interferers) $\lambda$ for $\alpha = 4$, $d = 1$, $W_\mathrm{p} = 1$ and different values of the constraints $W_\mathrm{max}$ and $\epsilon$.}
	\label{fig:T-vs-int}
\end{figure}

Next we will illustrate the analytic results just presented to get more insights on how the maximum link throughput $T^*$ is affected by the activity of the primary users, as well as the system constraints.
Fig. \ref{fig:T-vs-int} shows how the optimal constrained throughput $T^*$ in the smart meter-to-aggregator link behaves in relation to the density $\lambda$ of active mobile users that interfere in its communication.
As expected, increasing the density of interfering nodes decreases the maximum throughput achieved by the reference link, regardless of values of $W_\mathrm{max}$ and $\epsilon$ assumed.
It is interesting to see that $T^*$  exponentially decays with $\lambda$, which indicates that the secondary link performance is dramatically affected by the primary users' increase of activity.

We can also see from Fig. \ref{fig:T-vs-int} that the values of the constraints $W_\mathrm{max}$ and $\epsilon$ affect the maximum throughput curves.
Higher values of either $W_\mathrm{max}$ or $\epsilon$ leads to higher $T^*$, when  $\lambda$ is fixed.
While the result is intuitive for $W_\mathrm{max}$ (by increasing the transmission power, we obtain higher SIR and link throughput), it is not so when the outage constraint $\epsilon$ is considered: a less strict constraint leads to higher throughputs.

\begin{figure}[!t]
	\centering
	\includegraphics[width=0.7\columnwidth]{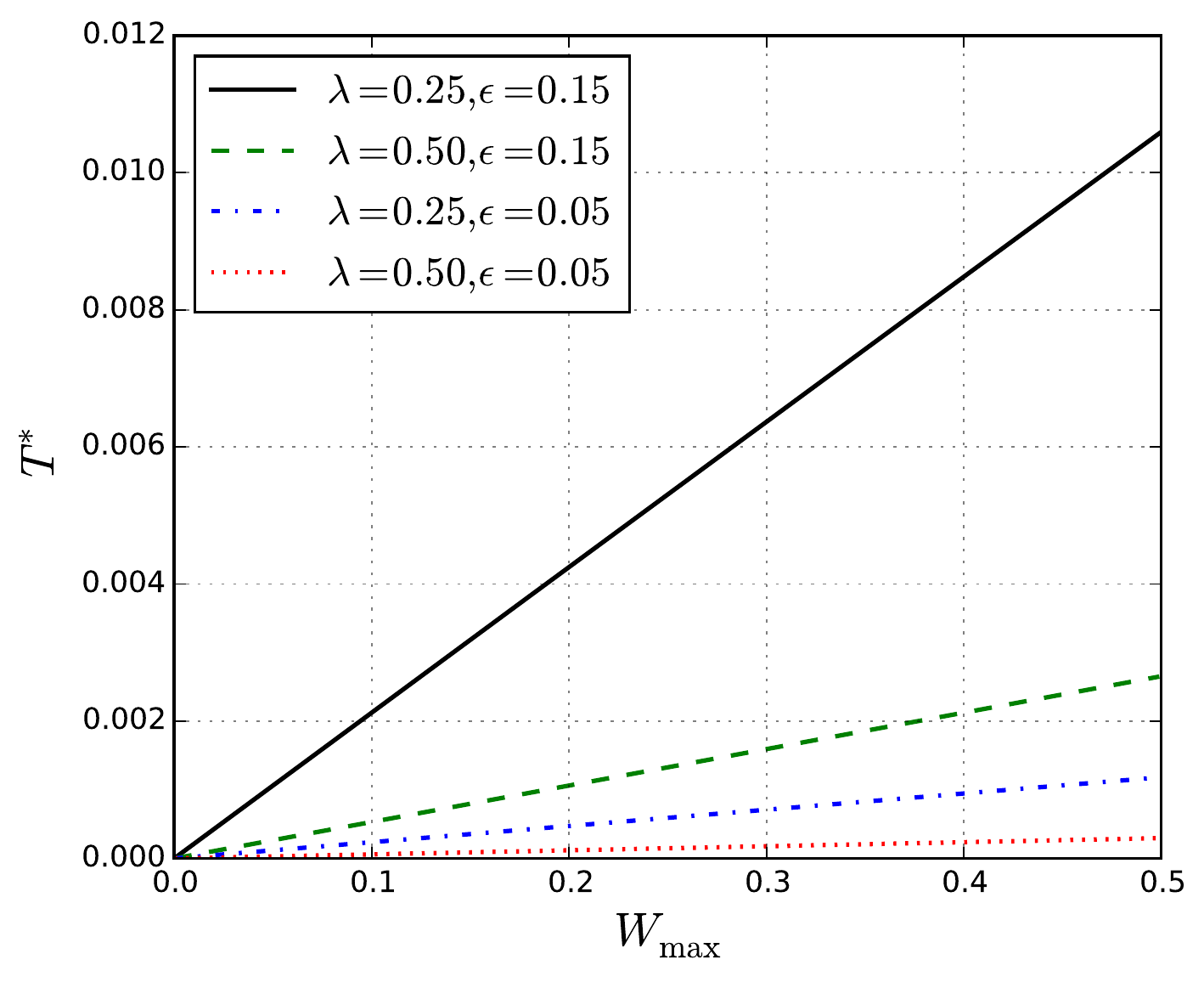}
	\caption{Maximum link throughput $T^*$ given in \eqref{eq:opt-T} as a function of the secondary power constraint $W_\mathrm{max}$ for $\alpha = 4$, $d = 1$, $W_\mathrm{p} = 1$ and different values of $\lambda$  and $\epsilon$.}
	\label{fig:T-vs-Wmax}
\end{figure}

\begin{figure}[!t]
	\centering
	\includegraphics[width=0.7\columnwidth]{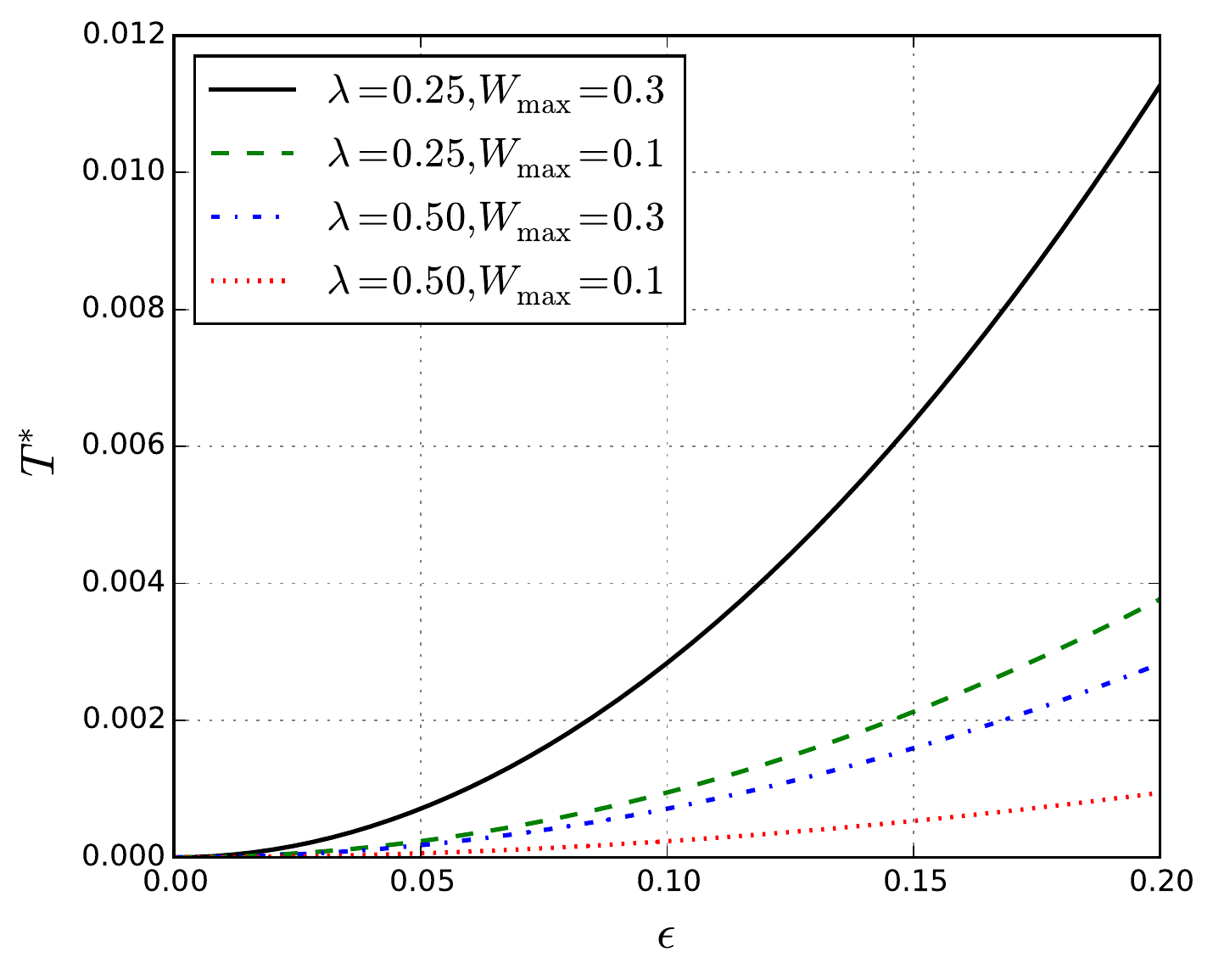}
	\caption{Maximum link throughput $T^*$ given in \eqref{eq:opt-T} as a function of the outage constraint $\epsilon$  for $\alpha = 4$, $d = 1$, $W_\mathrm{p} = 1$ and different values of $\lambda$ and $W_\mathrm{max}$.}
	\label{fig:T-vs-ep}
\end{figure}

To better understand those behaviors, we present in Figs. \ref{fig:T-vs-Wmax} and \ref{fig:T-vs-ep} the maximum throughput $T^*$ versus $W_\mathrm{max}$ and $\epsilon$, respectively.
Fig. \ref{fig:T-vs-Wmax} shows that the maximum throughput $T^*$  linearly grows  with $W_\mathrm{max}$, which was predicted by Corollary \ref{col:T-approx}.
This means that any relaxation in the power constraint $W_\mathrm{max}$ provides a linear gain in the secondary link throughput, whose slope is directly defined by the system variables; therefore, a combination between a relatively high outage constraint and low density of interferers provides the best performance.

The behavior of $T^*$ as a function of the outage constraint is more complicated since there is a trade-off involved, as shown in \eqref{eq:opt-T} and \eqref{eq:opt-T-approx}.
However, from our assumption that $\epsilon$ is a relatively small probability, i.e. $\epsilon \in (0,0.25)$, then $T^*$ is a (non-linear) crescent function of $\epsilon$.
Fig. \ref{fig:T-vs-ep} illustrates this growth, evincing that an increase $\Delta \epsilon$ for smaller values of $\epsilon$ leads to lower variation $\Delta T^* $ 
in the maximum throughput.
In this case, allowing for more outage events is more advantageous for the link: the spectral efficiency gains obtained by setting higher SIR thresholds dominates the system performance.
It is worth reinforcing that this behavior -- which is somehow counter-intuitive -- is only valid for lower values of $\epsilon \in (0,0.25)$.
If outages are not a constraint in our system, the results of Proposition \ref{prop:optimization-solution} should be reviewed under the perspective of Lemmas \ref{lemma:function-analysis} and \ref{lemma:optimal-beta}, and  Remark \ref{rem-beta-opt}.

\section{Outage events and signal reconstruction}
\label{sec:signal-rec}
We just showed in the previous section that allowing for more frequent outage events increases the link throughput.
However, outage events will affect the information that needs to be sent by the smart meter to the aggregator.

Let $x[n]$  be the discrete signal transmitted by the smart meter and  $\hat{x}[n]$ be the signal received by the aggregator, where $n=1,...,N$ with $N$ being the last sample.
We consider the $x[n]$ is the average power demand over a fixed period of time $\tau$.
The aggregator needs then to use $\hat{x}[n]$ to reconstruct the signal $x[n]$.
We assume here that the signal is reconstructed via linear interpolation between two adjacent points.
If the communication is perfect (i.e. $\hat{x}[n] = x[n]$) the interpolation is always between $\hat{x}[k]$ and $\hat{x}[k-1]$, with $k=2,...,N$.

This, however, is not the case in our model since outage events may occur due to the primary users activity.
If a sample is lost, the aggregator will interpolate the missing value(s) using the latest two received samples.
Consider the transmitted sequence: $x[k-2], x[k-1], x[k]$  with $k=2,...,N$.
If the samples $x[k-2]$ and  $x[k]$ are successfully received but $x[k-1]$ is not, the reconstruction is based on the linear interpolation of $\hat{x}[k] = x[k]$ and $\hat{x}[k-2] = x[k-2]$.
The estimation of the missing point is then $\hat{x}[k-1] = (x[k] + x[k-2])/2$.

To carry out this procedure, we use the ``The Reference Energy Disaggregation Data Set'' database \cite{kolter2011redd,REDD} from where we build our signal $x[n]$, which is the 15-minute average power demand over a timespan of $24$ hours (one day).
The information from the smart meter is transmitted to the aggregator every $15$ minutes (yielding $\tau=0.25$ hour and $N=96$ samples), which reconstructs the signal as previously described.

\begin{figure}[!t]
	\centering
	\includegraphics[width=\columnwidth]{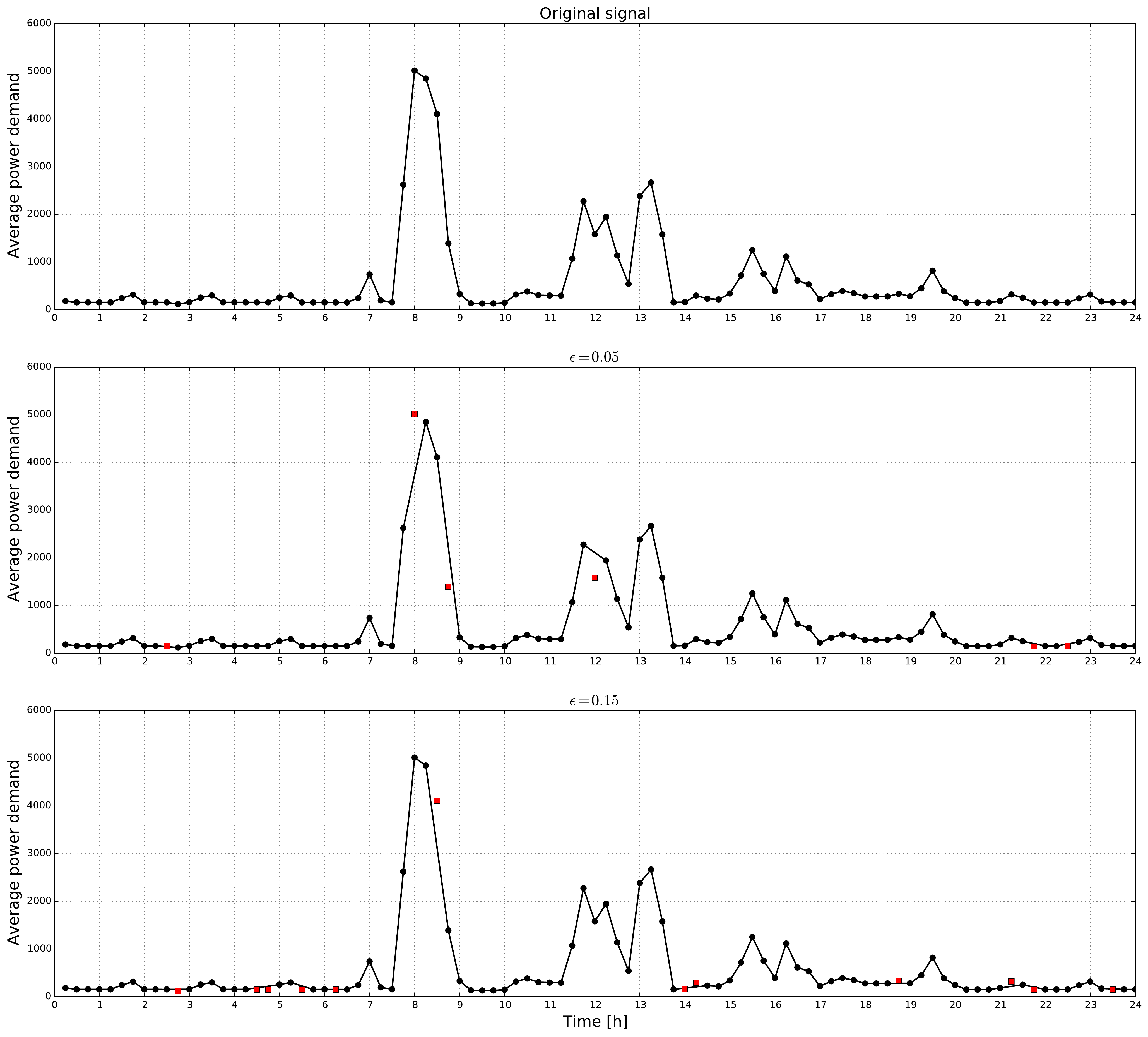}
	\caption{Average power demand of a house measured in watts during a period of 24 hours.  The signal is reconstructed in the aggregator as a linear interpolation between two subsequent points. If an outage happens, the point related to the power demand at that time is lost. On the top: perfect reconstruction. At the center: outage probability $\epsilon = 0.05$. On the bottom: outage probability $\epsilon = 0.15$. The red squares in the last two plots are the missing points (samples).}
	\label{fig:daily-consuption-example}
\end{figure} 

Fig. \ref{fig:daily-consuption-example} illustrates the effects of the outage probability on the signal reconstruction by showing the original signal and a snapshot of outputs assuming the constraints $\epsilon=0.05$ and $\epsilon=0.15$, where the missing points in the $\epsilon = 0.05$ and $\epsilon = 0.15$ curves are identified by the red squares.
At first sight, one cannot notice a big difference between the three signals, and the most noticeable differences are in the $\epsilon = 0.05$ curve (specifically in the ``peaks''), not in the $\epsilon = 0.15$ one.
This interesting fact happens due to the nature of the signal itself and the randomness of outage events.
The power demand signal seems to have a burst nature with a floor level and few peaks, which is related to personal habits when the house is occupied and people are engaged in activities like cooking or showering.

For example, we can see in Fig. \ref{fig:daily-consuption-example} that the peaks are around 8:00 in the morning and noon.
These peaks are probably related to people getting ready to work and having lunch.
Other than this, specially between 10:00 in the evening and 6:00 in the morning of the following day, we can see that the energy consumption is quite low and constant, most probably related to appliances in stand-by and refrigerator cycles \cite{Pipattanasomporn2014}. 
In this way, most of the samples $x[k]$ will have similar values.
If independent and identically distributed erasures occur, the probability that the estimated point approximates the missing one is high.

Consider again the transmitted sequence: $x[k-2], x[k-1], x[k]$ with $k=2,...,N$ and that the samples $x[k-2]$ and $x[k]$ are successfully received but $x[k-1]$ is not.
Then, the reconstruction is $\hat{x}[k-1] = (x[k] + x[k-2])/2$.
As in most of the cases $x[k] \approx x[k-1] \approx x[k-2]$, then $\hat{x}[k-1] = (x[k] + x[k-2])/2 \approx  x[k-1]$. 
However, during the peaks, this does not hold and errors become evident. 
The snapshot presented for $\epsilon=0.05$ is an unlucky one as far as the transmissions failure happened in the peak periods.
Conversely, although more samples were lost when $\epsilon=0.15$, they were mostly in the floor-level periods and the signal reconstruction was not affected in this specific snapshot.

To statistically analyze the effects of the outage events in the signal reconstruction, we use the root-mean-square deviation (RMSD) such that the reconstruction error of the $\hat{x}[n]$ is computed as:

\begin{equation}
\label{eq:rmsd}
\textup{RMSD} =  \sqrt{\dfrac{1}{N} \; \sum\limits_{k=1}^{N}(\hat{x}[k] - x[k])^2\;}.
\end{equation}

\begin{figure}[!t]
	\centering
	\includegraphics[width=0.7\columnwidth]{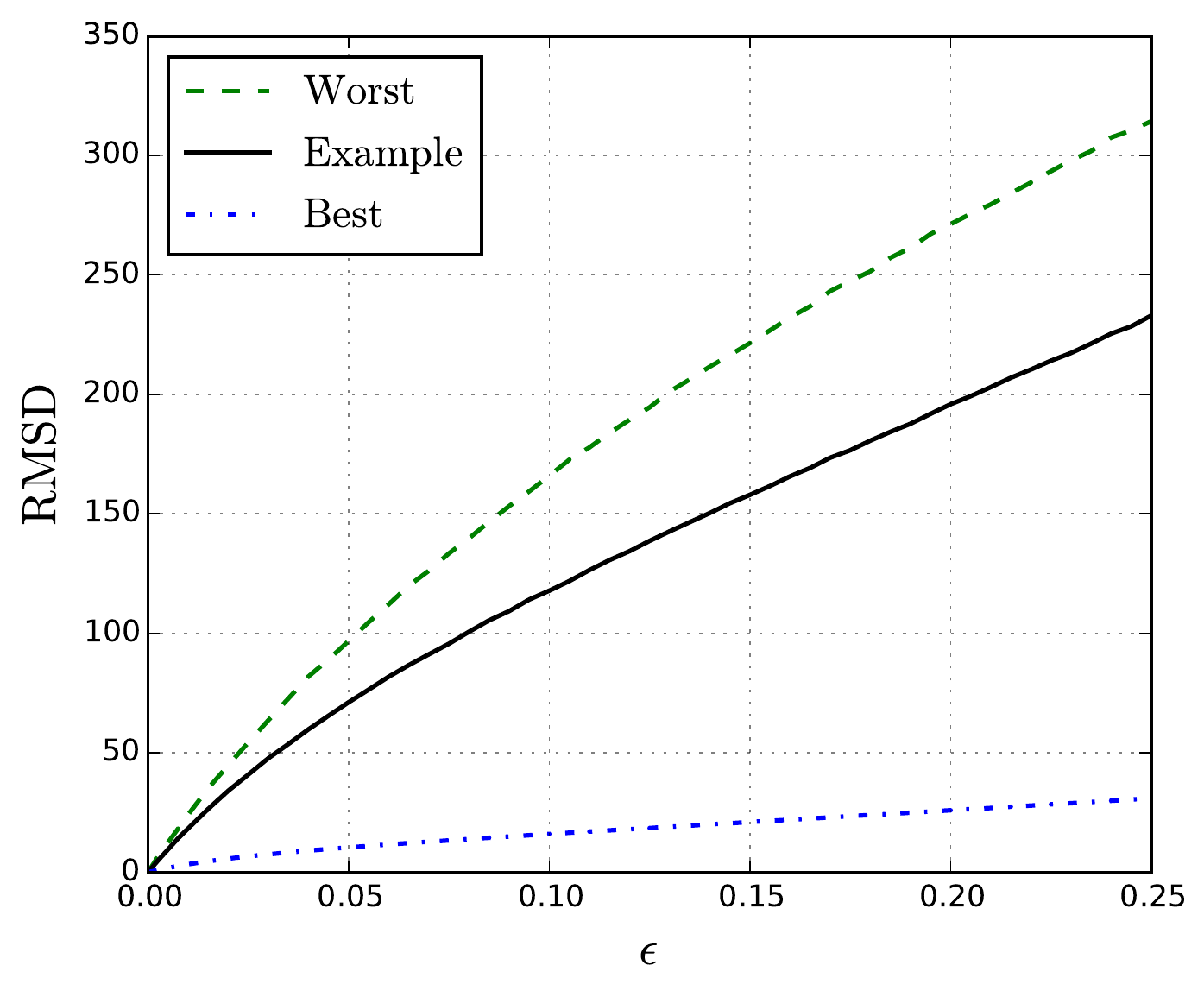}
	\caption{Root-mean-square deviation (RMSD) versus the outage constraint $\epsilon$. \new{The results are presented for the example given in Fig. \ref{fig:daily-consuption-example} together with the best and worst cases in terms of RMSD among $53$-household demand data from \cite{REDD}. The curves are the result of Monte Carlo simulations considering $10^5$ realizations for each point.}}
	\label{fig:outage-vs-rms}
\end{figure}

Fig. \ref{fig:outage-vs-rms} presents how the RMSD changes with the maximum allowed outage probability $\epsilon$.
As expected, more frequent outage events (indicated by higher $\epsilon$) leads to a worse signal reconstruction (indicated by higher RMSDs).
Our results show, however, that $\epsilon$ has relatively little effect on reconstructing the signal.
Let us first analyze the example presented in Fig. \ref{fig:daily-consuption-example} (black curve in Fig. \ref{fig:outage-vs-rms}).
For the highest value considered $\epsilon = 0.25$, the RMSD is $227$ watts for a signal with mean of $587$ watts and with range $\max x[n] - \min x[n] = 4895$ watts.
In this case, the normalized RMSD with respect to the mean is approximately $0.39$ while with respect to the range is less than $0.05$.
Looking back to the signal itself in top plot of Fig. \ref{fig:daily-consuption-example}, one can see that there are few points that, if erased, would cause a significant distortion on its reconstruction.

\begin{figure}[!t]
	\centering
	\includegraphics[width=0.7\columnwidth]{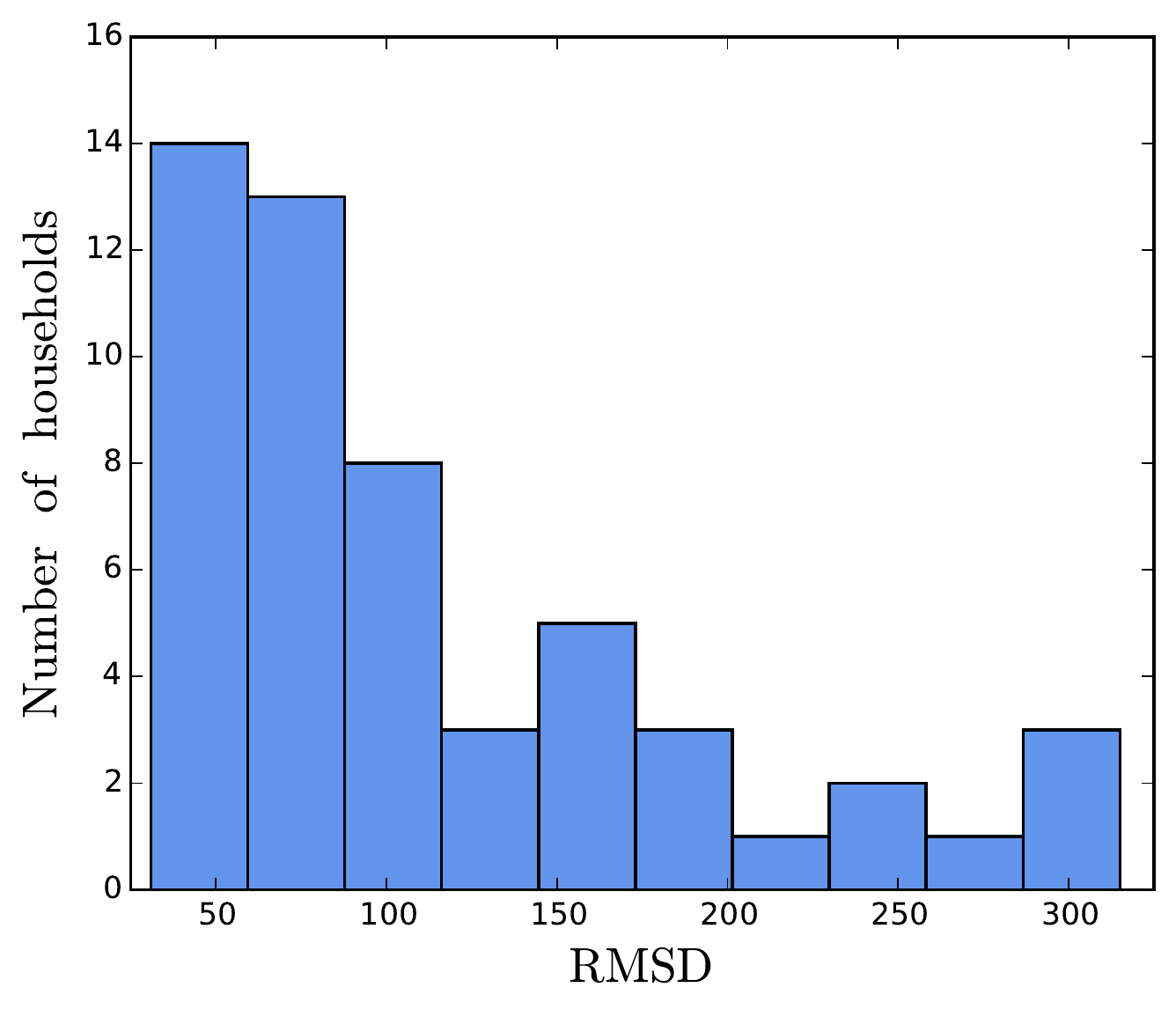}
	\caption{\new{Frequency diagram of the root-mean-square deviation (RMSD) considering data from $53$ households from \cite{REDD} for $\epsilon = 0.25$ versus the outage constraint $\epsilon$. The results are based on Monte Carlo simulations using $10^5$ realizations for each point.}}
	\label{fig:hist-rms-out25}
	\vspace{-3ex}
\end{figure}

\new{In order to have a more robust analysis, we also present in Fig. \ref{fig:outage-vs-rms} the best and the worst cases in terms of RMSD among $53$ households, whose data is available in \cite{REDD}.
For this scenario, our example is closer to the worst case than to the best one.
We further show in Fig. \ref{fig:hist-rms-out25} the frequency diagram (histogram) of RMSD for the scenario where $\epsilon = 0.25$.
As one can see, most households perform better than the example presented Fig. \ref{fig:daily-consuption-example} (RMSD $= 227$ W); they are in fact much closer to the best case presented in Fig. \ref{fig:outage-vs-rms}.
All in all, these results reinforces even more our argument that, for the scenario under consideration, the signal reconstruction is weakly affected by relatively frequent outage events.}

\section{Discussions}
\label{sec:disc}
Throughout the last two sections, we studied a spectrum sharing scenario where  a given smart meter-aggregator pair communicates using a licensed cellular uplink channel.
For this scenario, we analytically assessed how the system constraints for being a secondary user affect the link throughput and the signal reconstruction.
In this section, we start from the presented results to discuss how the proposed spectrum-sharing scheme could be implemented in actual systems.

\subsection{Relation between licensed and unlicensed users}
In our theoretical model, we assume that the secondary, unlicensed, users do not affect the primary users (base-stations).
As discussed before, this can indeed be the case when deploying our strategy in a real system.
Since the positions of the smart meters, aggregators and base-stations are fixed, the use of highly directional antennas in the smart meter-aggregator link with low transmit power will decrease the interference level at the base-station, which in turn still has capabilities of dealing with such residual interference.

Looking at the interference caused by the mobile users in the aggregator, our analysis assume that the density of these nodes are fixed. 
This, however, will probably not hold because the primary user activity probably would  intensify during periods that match with power demand peaks.
Then, in a practical scenario, this traffic variations is somehow predicted and the smart meter may adaptively change its operation point either by setting  a pre-determined, predicted, density of interferers to optimize its communication link (simpler solution) or by sensing and estimating such a density from time to time (more complex solution).

\new{Another important point that our model can be useful relates to Denial-of-Service (DoS) attacks in the physical layer \cite{lee2012physical}.
In this scenario, a node or a group of nodes intentionally interfering in the aggregator reception could characterize a DoS.
From the analysis presented here, we could infer how frequent outage events need to be to affect the signal reconstruction.
A dedicate study would be also possible by modeling the attackers as another Poisson point process.}

\subsection{Outage constraint and link robustness}
While the power constraint is required by the secondary link to not interfere in the primary users, the outage constraint is set to guarantee a minimum robustness at the secondary link.
Conversely to what one would expect, our results show that allowing for more frequent outage events improves the link throughput due to the contradictory effects of the SIR requirement on the system performance (lower outage constraints lead to higher SIR constraints, which results in higher spectral efficiency, while it decreases the success probability).

The question that arises from this result is \textit{how robust against outage events  the secondary link should be?}
One can only reply this answer knowing the information that is sent to the aggregator.
Our example shows that, if the information to be sent is the average power demand, the signal reconstruction is possible even with relatively loose outage constraints.
While this happens due to the nature of the input signal as discussed in the previous section, higher outage probabilities might be not desirable for other kind of signals or if the aggregator should provide some kind of feedback to the smart meter change the power demand behavior, as in strategies of demand-side management \cite{strbac2008demand}.

\subsection{Power demand signal processing and transmission}
\label{subsec:sig+trans}
The signal presented in Fig. \ref{fig:daily-consuption-example} exemplifies a $15$-minute sampling interval of the power demand of a household.
The signal characteristic indicates that time-based sampling might not be the most efficient way of collecting and then send the data to the aggregator.
For instance, by using event-based sampling\cite{tsividis2010event,Simonov2015}, the communication link should be much more robust (i.e. lower values of the outage constraint) since there will be much less redundant data and therefore the loss of any sample will have a more dramatic effect on the signal reconstruction.

Although this is not the focus of the present paper, we would like to mention that there is a trade-off between the sampling strategies and communication.
A more efficient way of sampling leads to less points for reconstructing the original signal and vice-versa.
If this is the case, the transmission strategies  and then the outage constraint should be evaluated in combination with the sampling procedure.
In the scenario used here, the time-based sampling generate redundant information about the signal so that outage events do not have drastic effects on the signal reconstruction.

\new{For this new study, however, it is important to have a stochastic characterization of power demand signal as in \cite{munkhammar2014characterizing,Sajjad2015}.
By doing so, the signal processing framework would be generalized and adapted to different consumption patterns.
This would allow us to study ways to optimize different sampling strategies (time-based, event-based and hybrid) and
transmission system designs (coding rate, medium access protocol and retransmissions), while considering the different requirements as reconstruction error and information privacy.
Looking at the reconstruction procedure, more advanced strategies like machine learning would probably offer better options to reconstruct the data than linear interpolation, but at expense of higher computational costs \cite{abu2014machine}.
From this perspective, it would be worth evaluating the cost of the sampling-reconstruction and the benefits of link throughput. 

In this paper, however, we have chosen to employ a two-dimensional analysis so as to understand the relations between optimal throughput and the reconstruction error itself; our goal is provide the knowledge of what combination of requirements is possible to achieve. 
For example, an application that requires a very high reliability would imply in a lower throughput.
Consequently, a cost-benefit analysis involving those aspects would require more information about the application requirement.}

Another different scenario might consider (hybrid) Automatic Repeat Request (ARQ) strategies in order to enhance the communication link, which reduces outage events \cite{Nardelli2012b}. 
This scenario may also include MAC protocols in order to control re-transmissions as well as coordination among nodes. 
One recent example of a MAC protocol designed in a similar scenario was developed in \cite{AijazTVT2014}, where the authors introduced a centralized strategy that resorts to a specialized frame structure that support co-existence between a cognitive and primary networks. 
Along the same line, the idea to include a routing protocol that protects the primary users while meeting the utility requirements of smart grid network was proposed in \cite{AijazTSG2015}. 
Another scenario of interest deals with security and privacy of the transmitted data, since there is a trade-off between security and reliability as pointed out in \cite{AlvesTIFS2015}, where the authors resort to information theoretical tools in order to guarantee security already at the physical layer.
All in all, combining these more advanced communication techniques to decrease the outage events with a more efficient sampling and enhanced security at the physical layer would be an interesting next step for the present work.

\section{Conclusions}
\label{sec:Conc}
This paper assesses a spectrum sharing scenario where smart meters send periodic information to an aggregator over licensed cellular uplink channels.
We assume that the secondary link uses directional antennas with limited transmit power so its interference  in the primary users can be neglected.
Mobile primary users, on the other hand, interfere with the aggregator reception.
Modeling the interferers' spatial distribution as a Poisson point process, we analyzed the secondary link throughput, finding then its optimal value under power and outage constraints.

Our results show that relatively high outage constraints surprisingly improve the link throughput for the cases of interest, even though more samples will be lost.
\new{It is worth mentioning the fact that smart meter reading application with scheduled intervals requires nowadays a reliability of at least $98\%$ \cite[Table 3]{kuzlu2014communication}; this is a fairly high value if compared to our results.
In fact, the discrepancy between the actual requirement and our results was a surprise.}
We also show that, due to the burst nature of the power demand signal that is transmitted in the smart meter-aggregator link, outage events do not have a dramatic effect in the signal reconstruction in comparison to the perfect transmission.

We plan to study in future works how outage events will affect the signal reconstruction under different sampling strategies.
In this way, we plan to build a joint sampling-transmission technique that can improve the system efficiency, as discussed in Section \ref{subsec:sig+trans}.
We also expect to implement the ideas proposed here in an actual demonstration to verify the validity of our assumptions and proposed optimization strategy.

\bibliographystyle{IEEEtran}

%
\end{document}